\documentclass[journal=aamick,manuscript=article]{achemso}

\pdfoutput=1 

\usepackage{chemformula} 
\usepackage[T1]{fontenc} 
\usepackage{amsmath,amsfonts,amssymb}
\usepackage{graphicx} 
\usepackage{subcaption} 
\usepackage{multicol} 
\usepackage[separate-uncertainty = true, multi-part-units = single, list-units = single, range-units = single]{siunitx} 
\DeclareSIUnit\Molar{\textsc{m}}
\DeclareSIUnit\rpm{rpm}
\DeclareSIUnit\ppm{ppm}

\DeclareSIUnit\kbt{k_BT}
\usepackage{chemmacros}  

\usepackage{xcolor}
\usepackage{upgreek}    
\usepackage{gensymb}
\usepackage[noabbrev,nameinlink]{cleveref}
\usepackage{multirow}
\usepackage{makecell}
\usepackage{colortbl}
\usepackage{booktabs}
\usepackage{lineno}
\usepackage{siunitx}

\title{Shape-dependent direction reversal in anisotropic catalytic microswimmers}

\author{Solenn Riedel}
\affiliation[Leiden University]{Soft Matter Physics, Huygens-Kamerlingh Onnes Laboratory, Leiden University, PO Box 9504, 2300 RA Leiden, the Netherlands}

\author{Mengshi Wei}
\affiliation[Leiden University]{Soft Matter Physics, Huygens-Kamerlingh Onnes Laboratory, Leiden University, PO Box 9504, 2300 RA Leiden, the Netherlands}

\author{Daniela J. Kraft}
\affiliation[Leiden University]{Soft Matter Physics, Huygens-Kamerlingh Onnes Laboratory, Leiden University, PO Box 9504, 2300 RA Leiden, the Netherlands}
\email{kraft@physics.leidenuniv.nl}


\begin{document}

\begin{abstract}The propulsion direction of active particles is a key feature in self-propelled systems and depends on the propulsion mechanism and environmental conditions. Here, using 3D micro-printed catalytically active particles, we experimentally show that the propulsion direction can change with increasing fuel concentration when the active particle possesses an anisotropic shape. 
We find that discs, tori, and bent rods reverse their direction of motion with increasing hydrogen peroxide concentration—moving with their inert side forward at low concentrations and with their catalytic side forward at high concentrations. In contrast, spheres and straight rods do not exhibit this reversal. We observe that direction reversal is independent of the base material composition of the swimmer and its size, and only occurs for anisotropic particles where, due to their elongated shape, the location of the solute concentration maximum is unstable and can be shifted by substrate-induced confinements. Our measurements suggest that in addition a change in the platinum-catalyzed reaction of hydrogen peroxide occurs, to which particles with elongated shapes that induce sufficient confinement are more sensitive. 
\end{abstract}

\bigskip

\section*{Introduction}
Catalytic active colloids are out-of-equilibrium synthetic systems that self-propel as a result of asymmetric solute gradients generated on the swimmer's surface \cite{Howse2007,Golestanian2005, Ebbens2014,Brown2014,Safdar2015}. Such chemically driven microswimmers are typically Janus particles \cite{Campbell2019} consisting of an inert and a catalytic side which are obtained by half-coating the particle with a metal catalyst. When dispersed in a fuel solution, the local decomposition of the fuel on the catalytic side generates non-uniform chemical fields around the particle that drive the motion of the swimmer \cite{Ebbens2014}. Speed \cite{Manjare2014}, directionality \cite{Lyu2022,Paxton2004,Xiao2025}, and swimming modes \cite{VanBaalen2023} are important features of this autonomous propulsion and the key to controlling these systems. 
\\ 
\\
The direction of propulsion of these Janus spheres is set by the location of the symmetry breaking metal coating and is parallel to the symmetry broken axis. Depending on the material of the catalytic cap the spheres either propel towards or away from the catalytic metal coating: Janus-spheres half coated with a thin platinum (Pt) layer typically swim inert side forward \cite{Ebbens2011, Campbell2013}, while Janus-spheres with a copper (Cu) coating swim catalytic side forward due to an inversion of the surface zetapotential of the metal hemisphere by the release of Cu\textsuperscript{2+} ions \cite{Sharan2023}. In both cases, the respective direction of motion is constant at all fuel concentrations. 
\\
\\
Reversing the propulsion direction, especially of anisotropic swimmers, dramatically changes the particle-particle interactions and consequently the collective behavior of the active system \cite{Riedel2024}. Predicting and controlling the direction of motion is therefore crucial when designing self-propelling units. However, a change in the direction of motion typically requires a different makeup of the particle. In 3D printed microtori, for example, different directionalities of the particle velocities were achieved by changing the chemical composition of the catalytic cap \cite{Baker2019}. Direction reversal can similarly be induced by changes in the experimental conditions and surrounding fluid. For example, adjustments in the local pH can influence the direction in which a swimmer propels either by altering the dominant reaction that occurs at the catalytic cap \cite{Liu2024} or by reversing the surface zeta potential \cite{Tan2022}. Likewise, changes in the illumination intensity can reverse the swimming direction of photocatalytic micromotors by a shift in propulsion mechanism from self-diffusiophoresis to self-thermophoresis \cite{Tong2021}. The addition of a cationic surfactant has furthermore been shown to reverse the propulsion direction of Pt-Janus microspheres because it inverts the surface charge of the particle, thereby leading to a reversal of the ionic current \cite{Brown2014}. 
\\
\\
Surprisingly, in our previous work on shape-dependent clustering of active particles, we found that crescent-shaped Janus-colloids exhibited opposite propulsion directions at different fuel concentration  \cite{Riedel2024}. At low fuel concentrations the particles moved with their inert cavity forward while at high fuel concentrations they moved with their metal-coated rounded side forward. The reversal in propulsion direction observed in crescent-shaped particles did not stem from a different make-up of the particles but occurred only upon a change in fuel concentration. Moreover, a change of the environmental condition alone cannot explain why crescent-shaped particles would reverse.  The reason is that despite extensive experiments on Pt-half coated Janus spheres driven by a catalytic decomposition of H$_2$O$_2$, no reports exist of a similar direction reversal in spheres suggesting that shape must play a decisive part.
\\
\\
Indeed, the particle shape of catalytic swimmers has been theoretically predicted to allow tuning of the propulsion direction. For anisotropic particle geometries, shape-induced effects on the diffusion and confinement of solutes can influence the near-field concentration gradients around the particle, and thus the orientation of the phoretic propulsive forces generated by the active site \cite{Michelin2017}. However, changing the propulsion direction of particles with the same shape was predicted to necessitate altering their relative electrophoretic mobilities or activities, which requires modifying the chemical properties of either the fuel or the particle. 
\\
\\
To shine light on this novel form of direction reversal, we here investigate how the particle shape can influence the swimming direction by examining active spheres, discs, tori, bend rods (crescents), and straight rods at different fuel concentrations. While all particles swim with their inert side leading at low fuel concentrations, we report a reversal of the swimming direction when the fuel concentration is increased for discs, tori and crescents. We show that the fuel concentration at which direction reversal for these shapes is observed coincides with a rapid change in pH. We hypothesize that this pH change causes a shift in the reaction mechanism, and that shapes which are elongated along the propulsion direction react to this shift by reversing their propulsion as they can tilt with respect to the substrate thereby affecting the solute confinement. Our insights provide the first methodological experimental study of a shape-dependent direction reversal stimulated by a change in fuel concentration.  
\\
\\
\begin{figure*}
\centering
\includegraphics[width=1.0\textwidth]{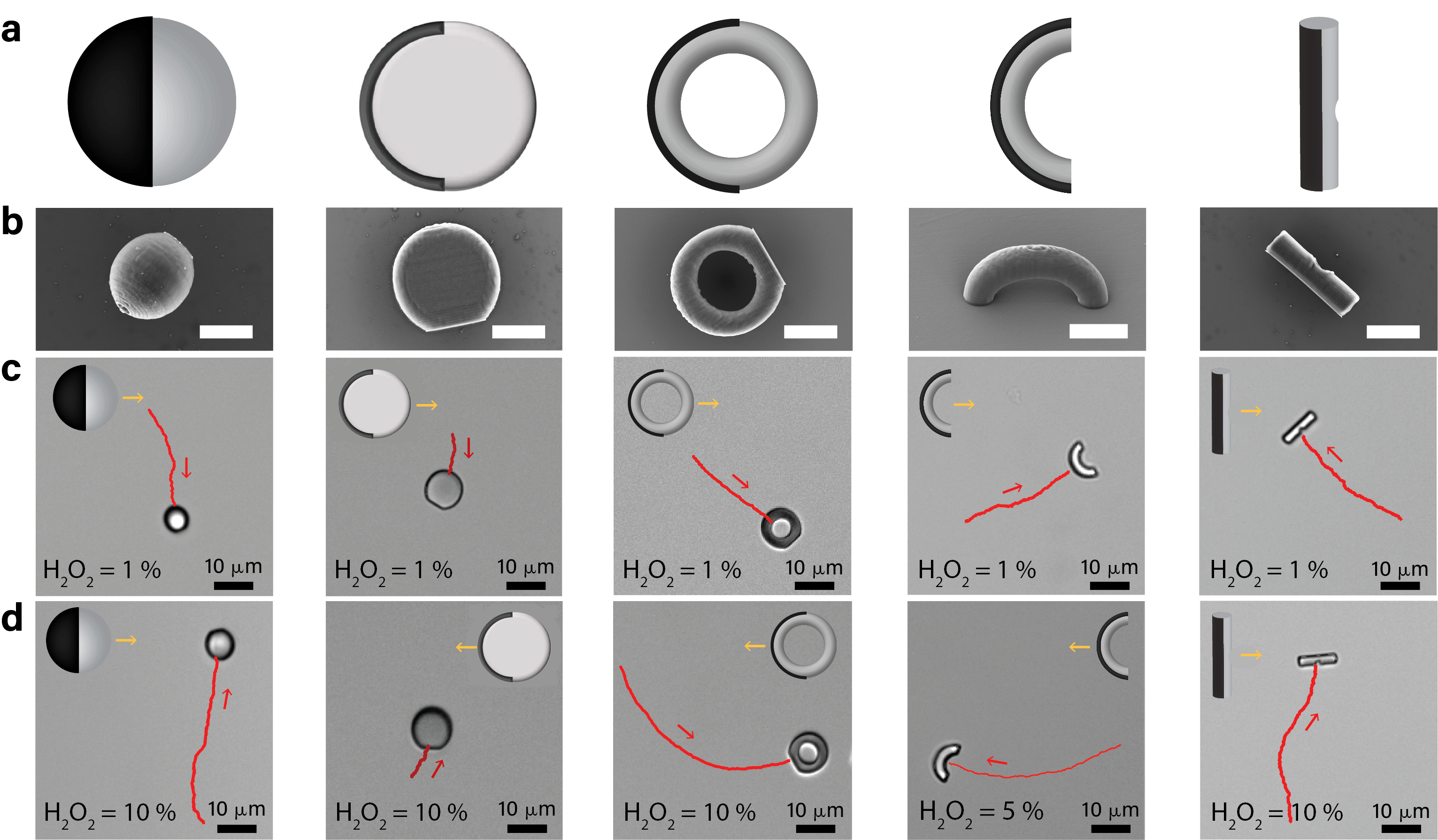}
\caption{\textbf{Shape- and fuel concentration dependent directionality of the propulsion direction.} (a) Schematics, (b)scanning electron microscopy (SEM) images  and (c-d) bright-field images with particle trajectory of 3D printed spheres, discs, tori, crescents and rods (left to right). The metal cap in the schematics is represented in black. The scale bar in the SEM images is 5 \si{\micro\meter}. 30s trajectory traveled by the active particle is shown in red. At low fuel concentrations (1 wt\% H$_2$O$_2$, c) all shapes swim inert-side forward. At higher fuel concentrations (10 wt\% H$_2$O$_2$, d) a reversal of the swimming direction is observed for discs, tori and crescents but not for spheres and rods. Spheres, discs and tori are cut off at the point where the particles were in contact with the substrate during 3D printing. This shape-defect allows to determine the particle' s inert side and thus its swimming direction. For rods a shape defect was intentionally introduced in form of a groove.}
\label{Fig_1}
\end{figure*}
\section*{Results and Discussion}
\textbf{Self-propelled 3D Microprinted Colloids.} To understand the influence of the particle geometry on the swimming direction, we compare the individual particle behavior of microswimmers that differ only in their shape. We make use of an efficient and reproducible 3D microprinting technique based on two-photon polymerization to obtain monodisperse anisotropic colloidal structures \cite{Doherty2020, Riedel2024}. We print particles of five different shapes keeping their maximal extension constant at 10$\mu$m: spheres, discs, tori, crescents and rods (see Fig.\ref{Fig_1} a and b). This choice of shapes was made because they are linked through simple changes in their symmetry and topology while covering a wide range of basic shapes. These include the widely employed spheres, for which no direction reversal with changing  H$_2$O$_2$ concentration has been reported thus far, and the crescents, for which we initially discovered the direction reversal. Spheres can be transformed into discs by reducing their z-dimension while tori are obtained from discs by introducing a cavity in the center of the structure. The torus structure is further halved to obtain a crescent (bend rod) and finally the crescent can be straightened into a rod, see the Materials and Methods section for more details on particle fabrication and particle dimensions, see SI for more details on the particle design. 
\\
\\
We render all particles active by sputter coating them with a 5 nm thick Pt/Pd (80/20) layer (cf. schematic in Fig.\ref{Fig_1} a) and dispersing them in aqueous hydrogen peroxide (H$_2$O$_2$) solution. Self-propulsion in this system of synthetic catalytic microswimmers is driven by solute gradients generated through the catalytic decomposition of H$_2$O$_2$ at the Pt/Pd cap~\cite{Howse2007, Ebbens2014}. For spheres, discs, and tori the propulsion direction can be identified by localizing a shape defect that originates from the contact area that the particles had with the substrate during printing. For rods, the inert side is recognizable by a small groove on the bottom side of the particle which we added for this purpose during particle design.
\\
\\ 
\textbf{Shape-dependent Swimming Direction and Direction Reversal.} We start by testing which of these shapes, beside the crescents, show direction reversal and disperse our 3D printed particles in both low and high fuel concentrations. In 1 wt\% H$_2$O$_2$ all particles move inert-side leading (Fig.\ref{Fig_1} c). However, when increasing the fuel concentration to 10 wt\% H$_2$O$_2$, we find that not only do crescents reverse their swimming direction, but also discs and tori now swim catalytic side forward, see Fig.\ref{Fig_1} d and Supplementary Videos 1-10. In line with literature, no direction reversal was observed for spheres and rods when the fuel concentration was changed. This suggests that the direction reversal is not caused by the material alone.
\\
\\
\begin{figure*}
\centering
\includegraphics[width=0.9\textwidth]{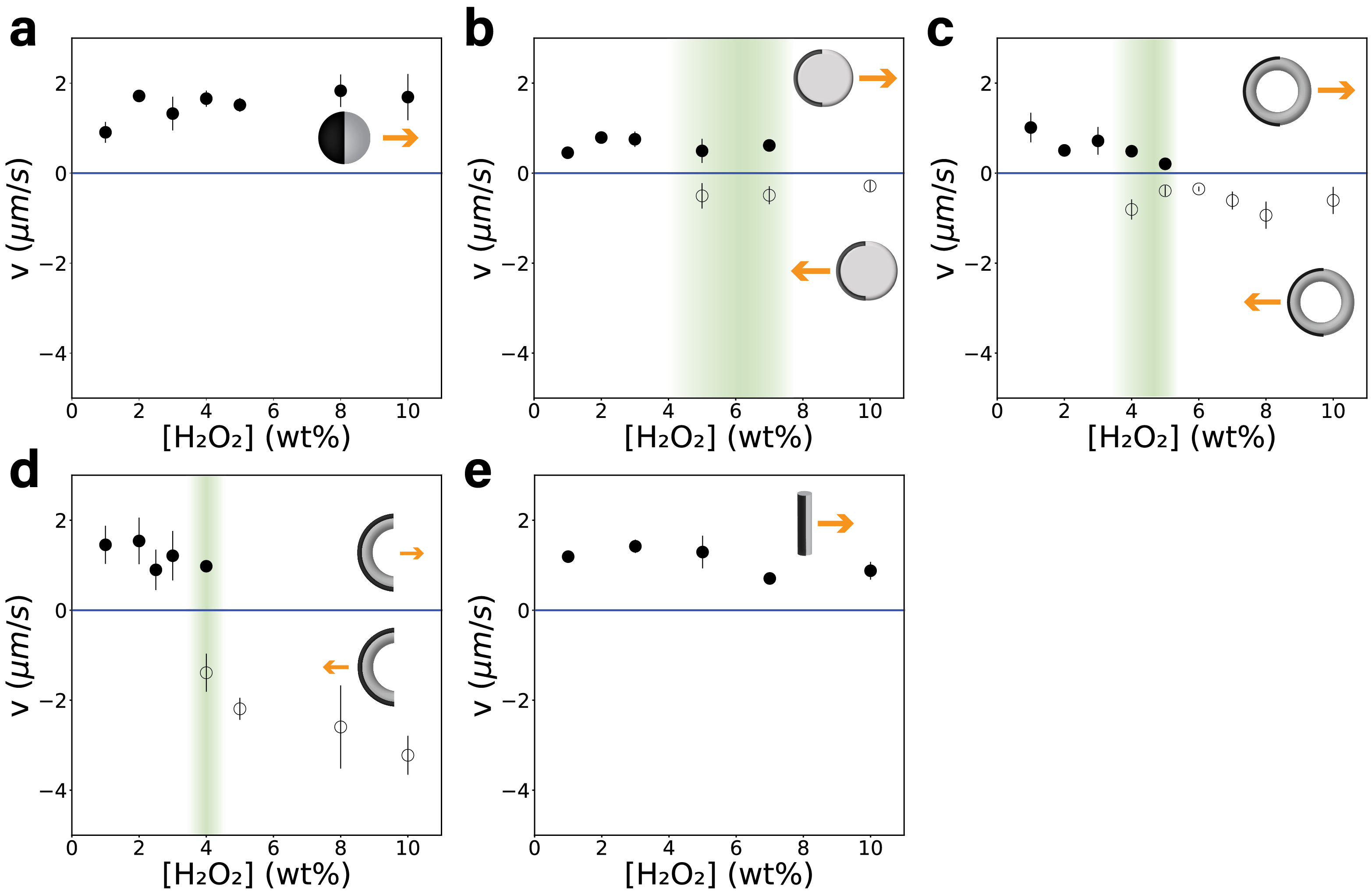}
\caption{\textbf{Influence of the fuel concentration.} Particle velocities were plotted over fuel concentrations for spheres (a), discs (b), tori (c), crescents (d) and rods (e). The transition region from positive velocity (inert-side leading) to negative velocity (catalytic-side leading) is highlighted in green. Plotted points are median values and error bars represent first and third quartiles.}
\label{Fig_2}
\end{figure*}
To better quantify this change in swimming direction, we sampled individual particle velocities at different fuel concentrations and identified the transition region where the reversal takes place, see Fig.\ref{Fig_2} (see SI for examples of particle trajectories). We mark the change in direction through a change in sign. Catalytic-side leading particles have a negative velocity while inert-side leading particles have a positive velocity. The speeds are in line with typical speeds reported for active spheres\cite{Ketzetzi2020}, crescents\cite{Riedel2024}, microtori\cite{Baker2019} and rods\cite{Vutukuri2016} with a catalytic platinum patch and of similar size. For all shapes where we observe two swimming directions the change in propulsion direction appears around 4-5 wt\% H$_2$O$_2$ followed by a transition region in which both directions are observed in significant amounts, i.e.  we report both species when the minority specie accounts for at least 30\% of the particles. 
This transition region becomes narrower the more the shape deviates from a sphere but straight rods swim again inert side-forward at all fuel concentrations. For discs this region stretches over a concentration range between 5-7 wt\% H$_2$O$_2$ while for tori it narrows down to a concentration of 4-5 wt\%. For crescents the transition occurs sharply at 4 wt\%. 
Thus, the swimmer's direction of motion is solely set by its shape and by the H$_2$O$_2$ concentration.
\\
\\
\begin{figure*}
    \centering
    \includegraphics[width=1.0\textwidth]{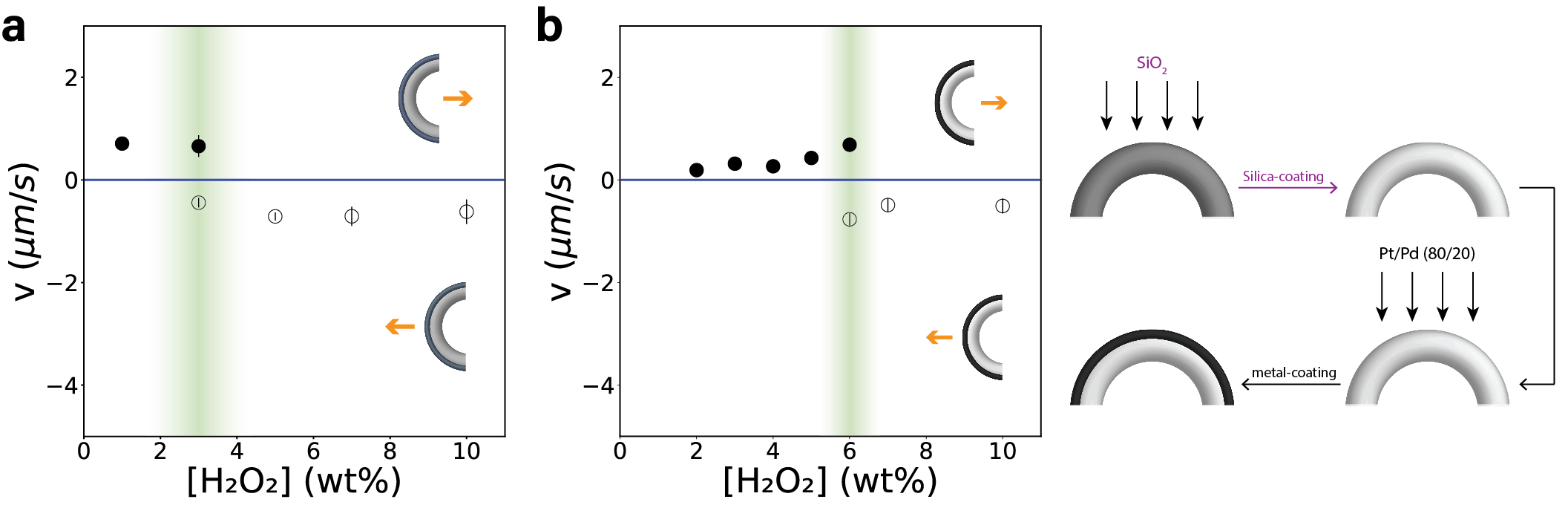}
    \caption{\textbf{Influence of the inert and catalytic surface material.} Dependence of the particle velocity on the fuel concentration for (a) crescents where a contamination of the catalytic cap material with PtO was excluded and (b) silica-coated crescents. The transition region, in green, from positive velocity (inert-side leading) to negative velocity (catalytic-side leading) shifts to lower fuel concentrations when the composition of the catalytic cap-material changes (absence of oxide). When the exposed inert surface is composed of silica the transition region shifts to higher fuel concentrations .}
    \label{Fig_3}
\end{figure*}
\textbf{Effect of the Material Composition of the Inert and Catalytic Sites.} As the cap material is known to influence the swimming direction of active spheres ~\cite{Sharan2023} and microtori~\cite{Baker2019}, we proceed to test whether the material of the inert and catalytic part contribute to the appearance of the direction reversal as well. We start by investigating the effect of the material composition of the catalytic site, and in particular investigate whether platinum oxide (PtO), which has been shown in SiO$_2$-Pt Janus microswimmers to change the swimming direction of particles~\cite{Lyu2022}, is present. We therefore measured whether our metal target showed traces of oxides by performing EDX measurements on the target side that gets exposed to the plasma during sputter coating. Indeed, contamination with PtO could be detected (see SI), however, the same target was used for all particles which only showed reversal in different fuel conditions. 
\\
\\
To further ensure that the propulsion direction of our anisotropic active particles is not linked to this contamination, we subsequently prepared swimmers with a Pt/Pd cap where the presence of PtO in the target material was excluded (see SI). We chose to perform material modifications on active crescents because these particles had shown the sharpest transition from positive to negative velocity and thus we expected the effect to be strongest for this shape. When PtO is absent in the catalytic cap-material, both directions of motion are still observed, the inert-side leading (positive velocity) at low fuel and the catalytic-side leading (negative velocity) at high fuel (Fig.\ref{Fig_2}a). However, the transition region from positive to negative velocity shifts to lower H$_2$O$_2$ concentrations for the Pt/Pd coating (Fig. \ref{Fig_3}a), demonstrating that the type of metal affects the precise conditions at which direction reversal occurs but not whether it occurs at all.
\\
\\
To complete the picture, we also varied the material composition of the inert surface of crescent-shaped microswimmers. For this, particles were fully coated with silica (SiO$_2$) using a high vacuum sputter coater with custom modifications, before applying the catalytic film on the convex side of the crescent, see Materials and Methods section. After dispersion in H$_2$O$_2$, the average propulsion speed of SiO$_2$-coated particles is found to be significantly lower across all fuel concentrations, see Fig \ref{Fig_3}b. This can be understood from the lower propulsion speeds for micromotors with a more hydrophilic surface, especially for larger sized particles \cite{Manjare2014}. Our 3D printed particles are made from a commercial photoresist (IP-Dip, Nanoscribe GmbH) that has a water contact angle (CA) of $\sim$\ang{50} (see SI). For bare silica particles, however, the CA is less than \ang{10} \cite{Kulkarni2008}. The surface of the SiO$_2$-coated particles is therefore more hydrophilic which explains the lower particle speeds. Still, just like for the IP-Dip particles presented in Fig.\ref{Fig_1}, we observe both swimming directions for SiO$_2$-modified crescents. The transition region shifts to higher fuel concentrations and appears at a somewhat higher H$_2$O$_2$ concentrations of 6wt\% H$_2$O$_2$, see Fig \ref{Fig_3}b.
\\
\\
We finally tested whether exposure to high fuel conditions might change the surface chemistry in an irreversible manner by exposing crescent-shaped particle first to high hydrogen peroxide concentrations before testing their swimming and direction reversal at lower fuel concentration. No change was observed, indicating the absence of any lasting changes in the surface chemistry. 
\\
\\
We showed that the shape-dependent direction reversal observed for crescent-swimmers is robust with respect to the material composition of the inert and catalytic side, only changing the H$_2$O$_2$ concentration at which the transition occurs. While contamination with PtO is not at the origin of the direction reversal, its presence enhances the velocity of the swimmer and delays the onset of the reversal. In contrast, a more hydrophilic inert surface reduces the crescent velocity while also shifting the transition region to a higher H$_2$O$_2$ concentration. There is hence no simple correlation between the particle speed and the fuel concentration at which direction reversal occurs.
\\
\\
\textbf{Effect of the Particle Orientation} At fuel concentration close to the transition region, we occasionally observe SiO$_2$-coated crescents to flip, sometimes combined with a change in swimming direction (see image sequence in SI).  The flipping suggests that the orientation of SiO$_2$-coated crescents in the xy-plane is not fully stable. Active spheres, in contrast, show a strong coupling of their symmetry-broken axis to the substrate leading to a small and stable tilt towards the substrate that slightly increases with fuel concentration \cite{Liu2025}. This has been proposed to arise from a balance between an activity-induced torque ($T_a$) and a gravitational torque stemming from the mass anisotropy due to the metal half-coating ($T_g$). The unstable orientation of our particles thus might provide a clue about the origin of the direction reversal mechanism. 
\\
\\
While tilting of spherical - and likewise our rod-shaped -  particles along thier propulsion direction can be executed without any steric constraints and does not impose any changes to the distances between the particle surface and the substrate (see Fig.\ref{Fig_4}a), this is different for anisotropic particles that do not posses a spherically symmetric cross-section along the propulsion direction ((see Fig.\ref{Fig_4}b). For discs, tori or bend rods (crescents),  tilting would induce an additional symmetry breaking in the propulsion direction. Moreover, the distance with respect to the substrate varies along those particles when tilted, and also changes with different tilt angles $\alpha$, see Fig.\ref{Fig_4}b. 
\begin{figure*}
    \centering
    \includegraphics[width=1.0\textwidth]{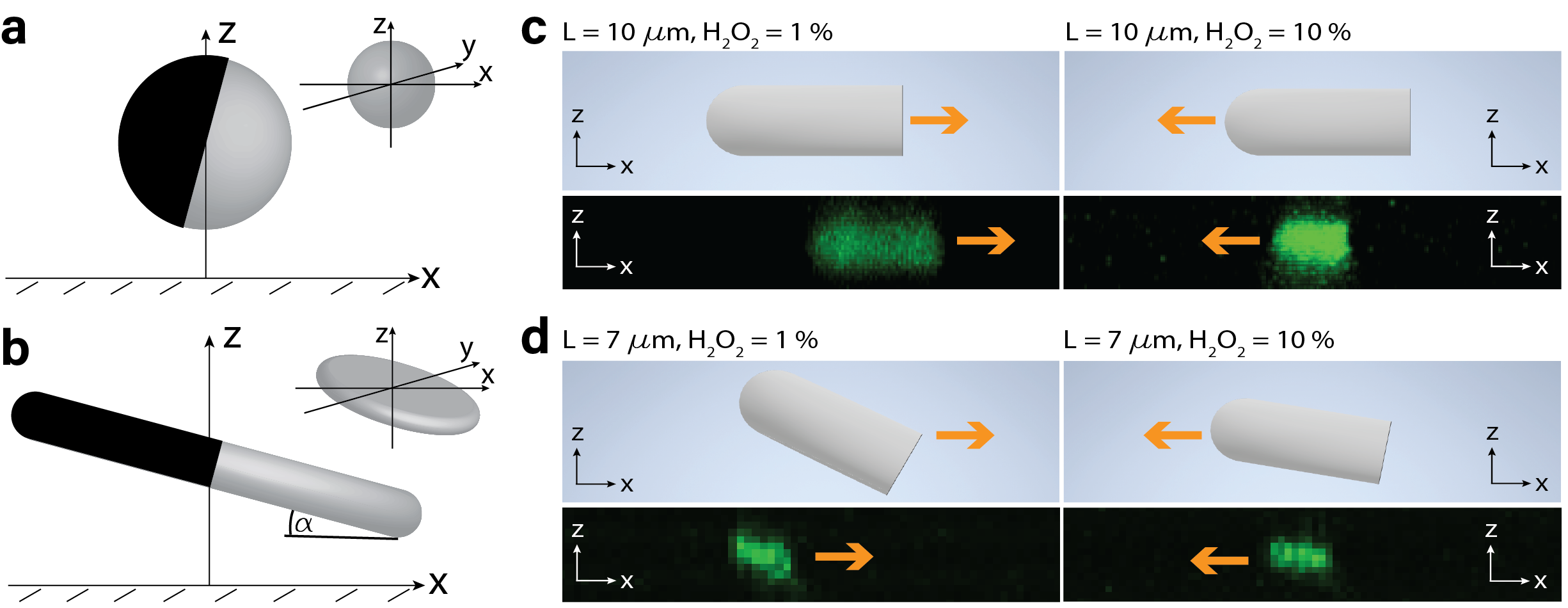}
    \caption{\textbf{Influence of a tilted particle orientation.} (a) A sphere tilted with respect to the substrate does not change the distances between the particle and the substrate. This is also true for rod-shaped particles propelling with their velocity director pointing perpendicular to their long axis and their projection in the zx-plane is identical to that of the sphere. (b) A tilted orientation in anisotropic shapes like discs, tori and crescents leads to an asymmetric confinement between the particle and the wall. This is exemplarily shown for the simplest of these shapes, a disc. (c) Projection of a confocal z-stack in the zx-plane and schematic illustrating the crescent orientation, for 10 \si{\micro\meter} sized crescents in 1wt\% (left) and 10wt\% (right) H$_2$O$_2$, respectively. No significant tilt can be observed for 10 \si{\micro\meter} crescents. (d) Schematic and z-stack projection for 7 \si{\micro\meter} crescents in 1wt\% (left) and 10wt\% (right) H$_2$O$_2$. In 1wt\% H$_2$O$_2$ smaller crescents occasionally adopt a tilted orientation. In 10wt\% H$_2$O$_2$ the tilt is much less pronounced.}
    \label{Fig_4}
\end{figure*}
\\
\\
To determine whether our crescent-shaped microswimmers assume a tilted orientation, we use confocal microscopy and perform a z-stack imaging of the moving particles in 1wt\% and 10wt\% H$_2$O$_2$, respectively. We note that taking a stack of images in the z-direction might cause a distortion of the image because the moving particle displaces in the x-direction between the top and bottom images of the stack. From the particle speed we estimate this x-displacement to be about 500nm, or 10\% of the particle size in this direction. 
\\
\\
For particles with 10$\mu$m cross-sectional length, no significant tilt can be observed in either fuel concentration, although the lower boundary of the particle at 10wt\% H$_2$O$_2$ suggests a small upward tilt in the moving direction, see Fig.\ref{Fig_4}c (top). When the cross-sectional length of the crescent is reduced to 7 \si{\micro\meter}, we occasionally observe a downward tilt of the crescent legs with respect to the substrate in 1wt\% H$_2$O$_2$ (Fig.\ref{Fig_4}c, bottom left). In contrast to Janus spheres which exhibit strong orientational locking, the orientation of the smaller crescents is not stable and fluctuates between a parallel and a tilted position (see SI). The downward tilt of the legs in the moving direction is in line with the tilt direction of active spheres~\cite{Liu2025}, and not an imaging artifact as the distortion of the z-stacked image due to the motion of the swimmer is opposed to the observed tilt. Occasionally, we also observe a tilt in the direction perpendicular to the propulsion, with the legs not being equally close to the substrate (see SI). 
When suspended in 10wt\% H$_2$O$_2$, a tilt is  much less pronounced but still present in 7 \si{\micro\meter} crescents and now away from the substrate and moving direction, see Fig.\ref{Fig_4}c (bottom right). 
\\
\\
We attribute the difference in behavior between smaller and larger crescents to the effect of their size on the competition between activity-induced torques and the gravitational torques required to lift and tilt the anisotropic particle. In the absence of activity, Janus-crescents adopt a flat, parallel orientation relative to the substrate. For an active torque to induce a tilt not only must the mass asymmetry caused by the metal coating be overcome, but the particle also needs to be lifted away from the substrate to be able to tilt. The active torque achieves this for the smaller and lighter crescents reorienting them such that the director points into the substrate, but not for the larger heavier ones. At high fuel concentrations, the particles propel with their convex side forward, generating a propulsion force directed away from the legs. This reverses the direction of the activity-induced torque and consequently reduces the tilt.
\\
\begin{figure*}
    \centering
    \includegraphics[width=0.5\textwidth]{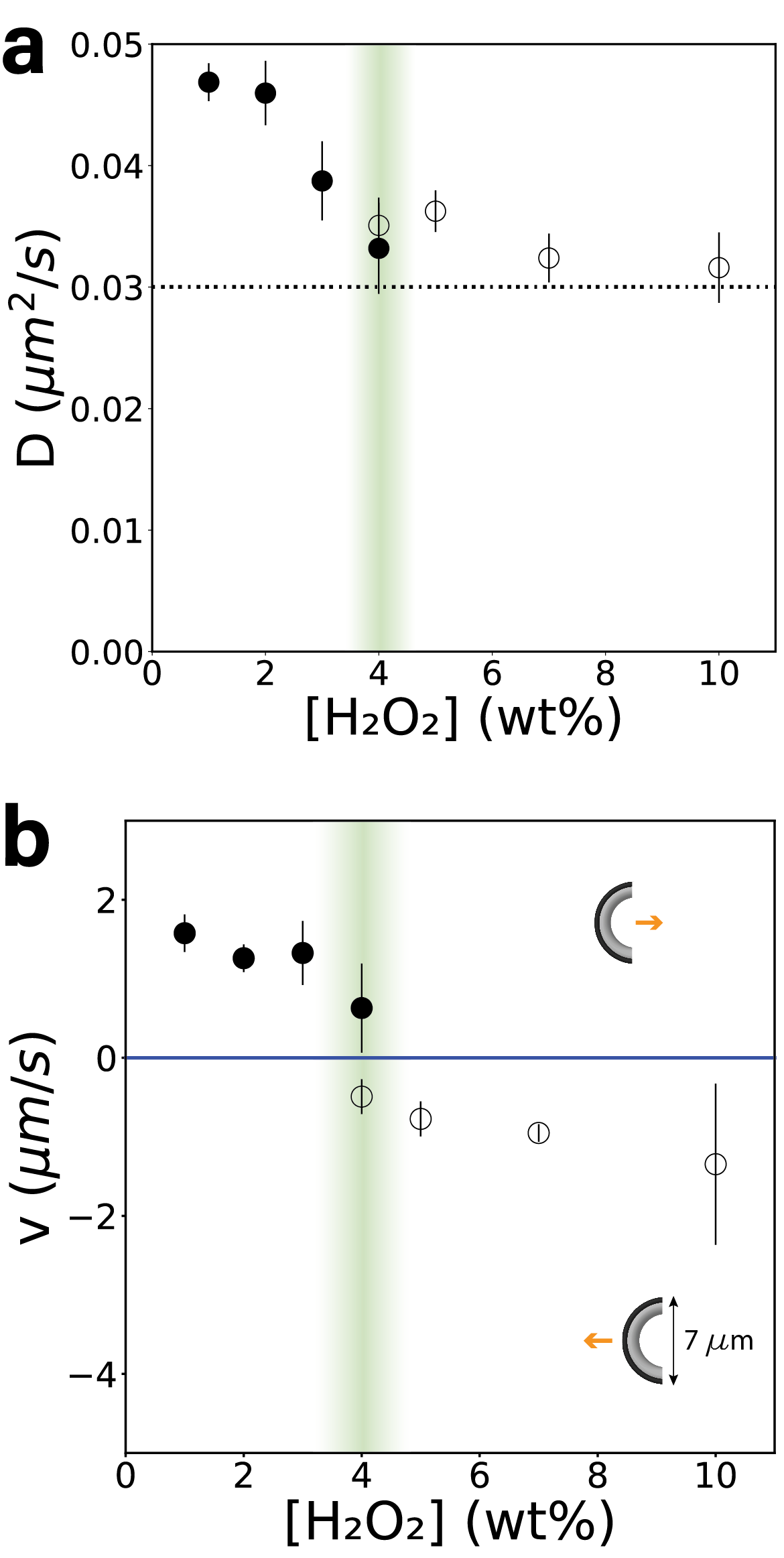}
    \caption{\textbf{Diffusion constant and particle velocity for 7 \si{\micro\meter} crescents in different fuel concentrations.} (a) Height-dependent diffusion constant $D$ for 7 \si{\micro\meter} crescents in different fuel concentrations. A stronger tilt for inert-side leading crescents in lower fuel reflects in a higher $D$. The dotted line indicates the expectation value for an equally sized sphere with a radius of 3.5 \si{\micro\meter} close to the substrate \cite{Ketzetzi2020}. (b) Dependence of the average particle velocity on the fuel concentration. The transition region, in green, is unaffected by a change in particle size.}
    \label{Fig_5}
\end{figure*}
\\
Further evidence for the presence of a tilt stems from an analysis of the diffusion coefficient (see SI for diffusion constants of other shapes). The stronger tilt orientation of inert-side leading 7 \si{\micro\meter} crescents should cause the center of mass of the particle to shift further away from the substrate. This increased height should be reflected in a higher diffusion coefficient $D$ \cite{Ketzetzi2020} and particle speed due to reduced hydrodynamic interactions with the substrate. 
Before the direction reversal, the average diffusion constant of 7 \si{\micro\meter} crescents is indeed significantly larger than after, suggesting that the particle has assumed a flatter orientation after direction reversal, see Fig.\ref{Fig_5}a. A larger height between the particle and the substrate also means that smaller particles experience a weaker hydrodynamic coupling with the wall which is expected to stabilize their orientation. This is in line with our observation that at fuel concentrations close to the transition region, some 7 \si{\micro\meter} crescents have also been observed to flip (see SI), which suggests an orientational instability.
\\
\\
The average particle velocities of 7$\mu$m crescents at low H$_2$O$_2$ concentrations is comparable to the values obtained for 10 \si{\micro\meter} crescents and the location of the transition region is unaffected by a change in particle size (cf. Fig.\ref{Fig_2}d). Beyond the transition the average particle velocity increases with increasing fuel concentration but does not reach the high values observed for 10 \si{\micro\meter} crescents (cf. Fig.\ref{Fig_2}d).
\\
\\
Based on these findings, we conclude that for 7 \si{\micro\meter} crescents the propulsion direction results in different particle orientations with respect to the substrate and that the swimming direction reverses sharply at 4 wt\%. Since larger crescents also show direction reversal with the same sharp transition region but are not significantly tilted with respect to the substrate, we conclude that a change in particle orientation is an effect of different propulsion directions and not the cause for it.
\\
\\ 
\textbf{Effect of the fuel pH} After having excluded possible direction reversing effects in connection with the active particle itself, like its material composition, its orientation or its size, we now turn towards investigating the changes occurring in the external experimental conditions when the fuel concentration is varied. For discs, tori and crescents the directionality of the particle velocity depends on the H$_2$O$_2$ concentration and starts reversing at about 4 wt\% fuel (cf. Fig. \ref{Fig_1}e). As H$_2$O$_2$ is a weak acid, the pH of the solution decreases with increasing H$_2$O$_2$ concentration for all fuel solutions (Fig. \ref{Fig_6}a). We note that the pH changes sharply in the concentration range where discs, tori and crescents propel inert-side forward while a more gradual change is observed in the range where these particles move catalytic-side forward. The change in pH also affects the bare particles' surface zeta potential, which decreases with increasing fuel concentration, see Fig. \ref{Fig_6}b. Because the surface zeta potential always remains negative (see Fig. \ref{Fig_6}b)  a reversal in phoretic mobility can be ruled out to be the cause for the direction reversal \cite{Heckel2019,Zhou2021,Tan2022}. 
\\
\\
\begin{figure*}
    \centering
    \includegraphics[width=1.0\textwidth]{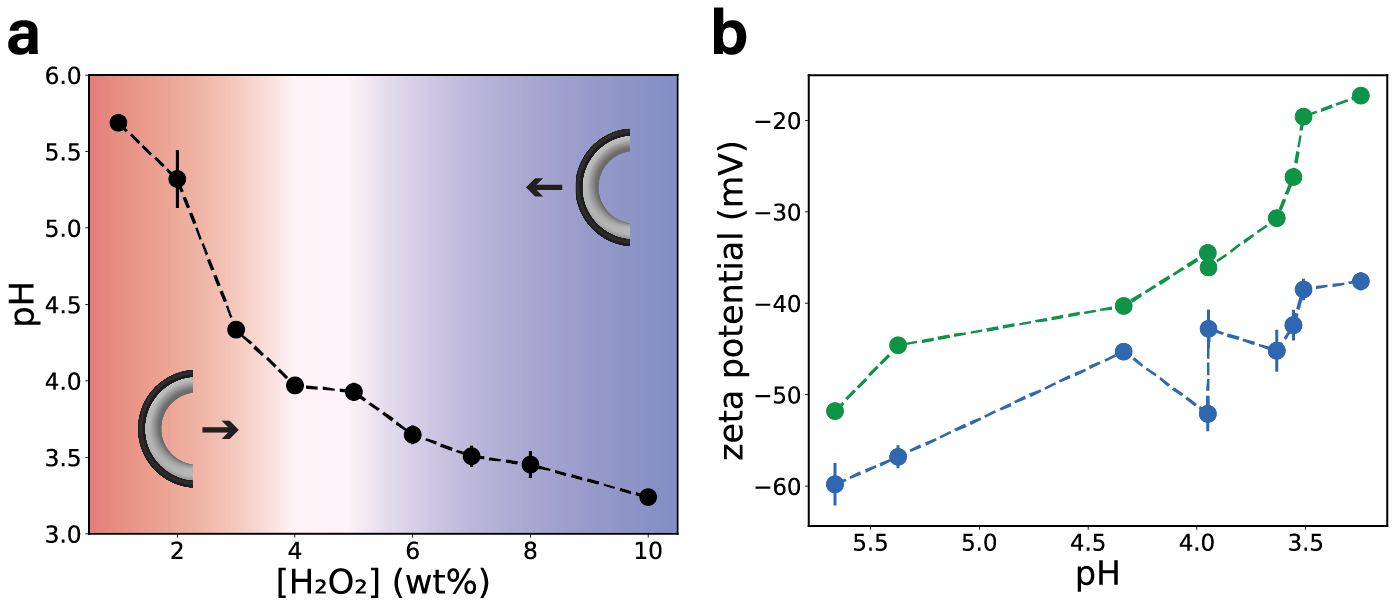}
    \caption{\textbf{The direction reversal occurs at a fuel concentration where the slope of the pH curve changes.} (a) H$_2$O$_2$ concentration of the fuel solutions plotted against pH. The concentration range where discs, tori and crescents propel inert-side forward is highlighted in red. The blue range corresponds to fuel concentrations where catalytic-side leading propulsion was observed. The transition region where both motion modes are observed is highlighted in white. Data points were connected by a dashed line to guide the eye. (b) The surface zeta potential increases with decreasing pH but always remains negative. Results are shown for particles made from droplets of IP-dip resin cured under UV irradiation for 2h  (top curve) and overnight (bottom curve), see Materials and Methods for details.}
    \label{Fig_6}
\end{figure*}
Having established that the direction reversal is not caused by environmental effects on the base material of the swimmer, we consider whether a change in pH has an effect on the reaction occurring at the catalytic cap. It has been shown that a change in local pH can alter the dominant reaction occurring at the catalytic cap and that this results in a reversal of the propulsion direction \cite{Liu2024}. In the catalytic decomposition of H$_2$O$_2$ by platinum nano-particles, two reaction pathways have been reported to compete, with the dominant reaction products being pH dependent \cite{Liu2014}. Under acidic conditions OH radicals form preferentially (peroxidase-like catalysis) while H$_2$O and O$_2$ are mainly produced under neutral and alkaline conditions (catalase-like catalysis). Since the production of OH radicals significantly increases between pH 6 and 1 and the pH of 1-10 wt\% H$_2$O$_2$ solutions lies in a range between pH 5.7 and 3.2, we hypothesize that the reaction mechanism for our anisotropic Janus particles shifts from a catalase-like catalysis towards a peroxidase-like catalysis with decreasing H$_2$O$_2$ concentration. A change in the catalytic reaction products would impact the near field concentrations and distribution of solutes and lead to altered flows. Even more so, it might change the mechanism of propulsion from one where charged species induce surface slip flows to one where neutral species take on that role.
\\
\\
\begin{figure*}
    \centering
    \includegraphics[width=1.0\textwidth]{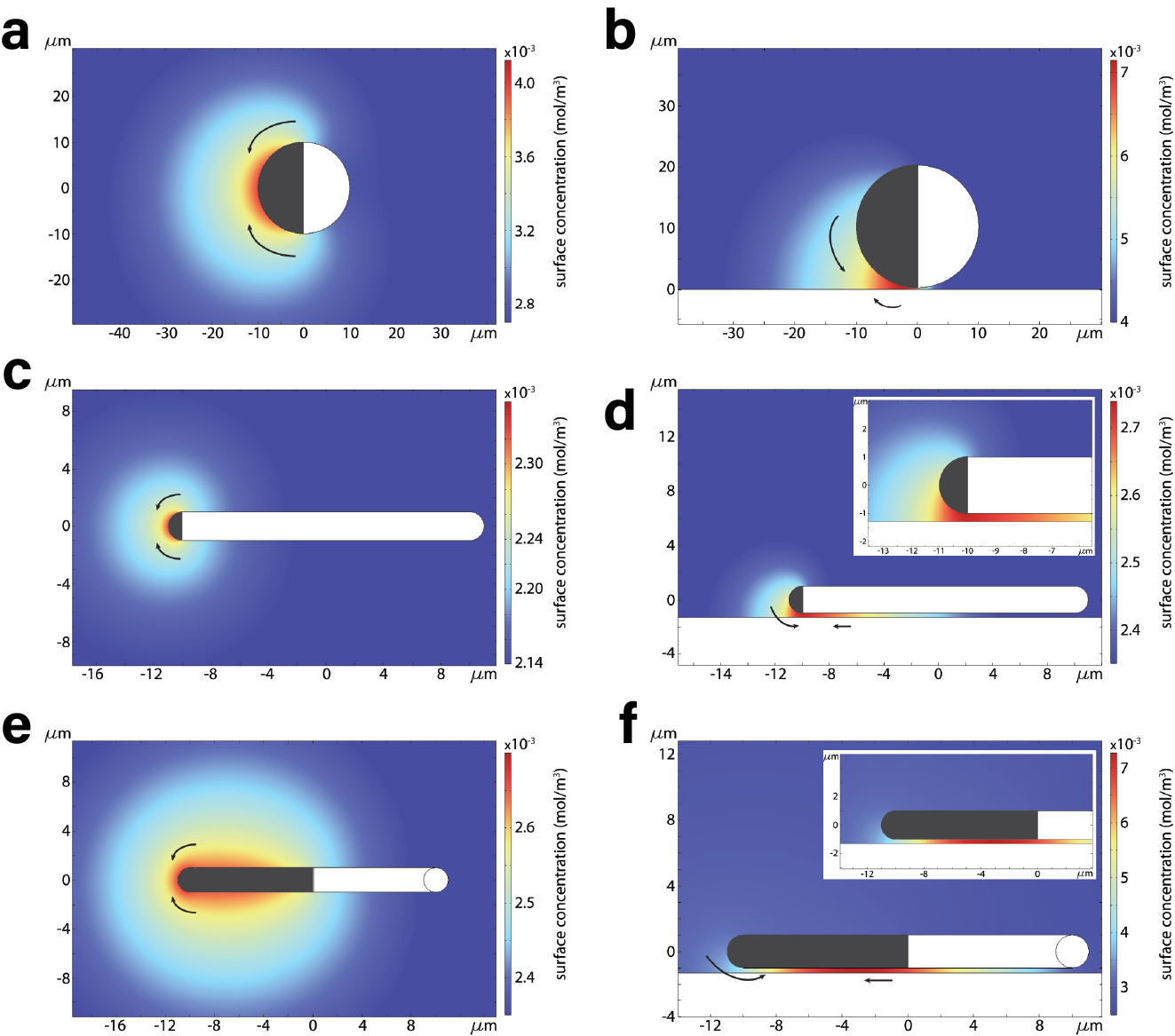}
    \caption{\textbf{COMSOL simulation of the spatial distribution of the chemical product $\mathrm{H}^+$ around Janus particles of different shapes.} Panels (a), (c), and (e) show the distribution in the bulk, while panels (b), (d), and (f) show the distribution near a substrate. (a)–(b) represent a Janus sphere with a hemispherical cap (black) producing $\mathrm{H}^+$ at a flux of $J_{0,\mathrm{H}^+}$. (c)–(d) show a Janus disk where only the curved edge (black) emits $\mathrm{H}^+$ with a flux of $J_{0,\mathrm{H}^+}$. (e)–(f) depict a Janus disk where the curved edge produces $\mathrm{H}^+$ with a flux of $J_{0,\mathrm{H}^+}$ and the flat half-side emits at a reduced flux of $\frac{1}{3}J_{0,\mathrm{H}^+}$, mimicking the inhomogeneous coating of the active layer. The surface slip flows are schematically indicated with black arrows. See Materials and Methods for further details.}
    \label{Fig_7}
\end{figure*}
\textbf{Shape-dependent Solute Confinement} A changing propulsion mechanism alone cannot explain why discs, tori, and bend rods reverse and spheres and rods do not. To rationalize this observation we note that besides the chemistry of the catalytic reaction, the swimmer's shape strongly influences the solute transport and thus decisively determines the concentration gradient profile around the particles ~\cite{Michelin2017}. A shape-dependent shift of the location of the maximum concentration can reverse the slip flows on the surface of the catalytic metal cap and consequently the direction of motion of the particle. Half-coated oblate particles similar in shape and coating to the discs employed here have been predicted to possess their highest solute concentrations neither at the highly curved metal-coated edge nor at the transition from metal to polymer, but in between~\cite{Michelin2017}. We hypothesize that the location of the concentration maximum in this type of concentration profile is unstable towards both directions and that it can shift more towards the edge of the particle or more towards the metal-polymer transition line. Using COMSOL simulations (cf. Materials and Methods section for more details), we show that additional confinement of the solutes due to the presence of the substrate significantly influences the concentration profile around the particle. For a half-coated Janus sphere, the solute concentration gradient typically has its maximum at the active pole (Fig.\ref{Fig_7}a). Introducing a nearby wall causes this maximum to shift closer to the wall, see Fig.\ref{Fig_7}b. In the case of an oblate particle with a catalytic cap covering only the rounded edge, the same confinement effect can push the concentration maximum underneath the particle, toward the center of the confined area (Fig.\ref{Fig_7}c and d). This shift becomes even more pronounced when the catalytic coating extends beyond the edge and partially covers both the top and bottom surfaces of the particle, see Fig.\ref{Fig_7}e and f. The flux of reaction products on the flat sides was intentionally reduced relative to the flux on the curved edge to mimic an inhomogeneous coating, as is likely the case for our anisotropic swimmers. We expect such confinement-induced changes to the concentration profile to be important enough to reverse the direction of the surface slip flows, thereby leading to a reversal in the propulsion direction.
\\
\\
To consider whether shape-induced confinements play a role here, we start by noting that the particles used in this work were all sputter coated by Pt from the top. This leads to a gradient in the Pt coating thickness on the particle, with highest thickness and thus highest solute concentrations at the highest points. This should provide a propulsion force away from the platinum, for all shapes. When viewed from the side the anisotropic shapes of bend rods, tori, and discs should have a profile similar to the oblate particles with inhomogeneous coating mentioned above (cf Fig.\ref{Fig_7}e and f), which means that the location of their concentration maximum is most probably also unstable. For stronger concentration gradients of solutes at higher fuel concentrations, we expect the additionally induced confinement between the particle and substrate to significantly influence the location of the maximum concentration and to shift it towards the middle of the confined area. This then results in a reversal of the gradient profile and consequently in the reversal of the propulsion direction, i.e. towards the platinum. 
\\
\\
More evidence for the important role of shape-dependent solute confinement stems from the width of the transition region between forward and backward swimming. The transition region is the broadest for discs, narrowest for microtori and becomes very sharp for crescents, in line with the increasingly smaller confinement area that their particle shape induces. Moreover, we expect a positive feedback mechanism once the swimmer moves into a certain direction. A motion towards the cap would drag the solute gradient further underneath the particle while a propulsion away from the cap would drag it closer to the edge of the particle. This could explain the presence of both species in the transition region that we have observed. Additionally, such a shape-dependent confinement should be independent of the precise material of the particle, in line with our experiments above. 
\\
From these considerations we attribute the reversal in the propulsion direction observed for discs, tori, and crescents to their anisotropic shape, which allows for solute confinement and tilting, and which is making these particles more sensitive to a pH-dependent change in the reaction mechanism.
\section*{Conclusion} In summary, we report a reversal of the propulsion direction with increasing fuel concentration for catalytically active anisotropic Janus colloids half-coated with Pt/Pd and suspended in aqueous H$_2$O$_2$ solution. This direction reversal is shape dependent and can be observed for discs, tori, and bend rods (crescents), but not for spheres and straight rods propelling along their short axis. At low fuel concentrations, particles propel with their inert-side forward while at higher fuel concentrations the swimming direction changes and particles move catalytic-side forward. In contrast, spheres and rods always move inert-side leading independent of the H$_2$O$_2$ concentration. We find that the fuel concentration at which the directionality of the particle velocity switches from inert-side leading to catalytic-side leading, is influenced by the material of the microswimmer and by the presence of oxide-contamination in the catalytic Pt-cap. However, the occurrence of the reversal itself is not tied to the material. Instead, based on the change in pH measured for the range of H$_2$O$_2$ solutions, we propose that a shift in the dominant propulsion mechanism changes the near field concentration distribution around the particle, thereby affecting the phoretic mobilities which leads to a reversal in propulsion direction of the active particle~\cite{Liu2014}. However, as the propulsion direction of spheres and rods is unaffected by the change in pH, we hypothesize that the anisotropic shape of the swimmer is crucial to obtain a reversal. 
\\        
\\
Specifically, we expect the particle's geometry to influence the location of the solute concentration maximum, which for discs, tori, and crescents is sensitive to additional solute confinement near the substrate. This shape-dependent solute confinement becomes especially significant with stronger gradients, where it leads to a shift of the concentration maximum along the particle's surface thereby reversing the concentration gradient profile and thus the surface slip flows and motion direction. Our results suggest that the shape-dependent confinement area needs to be large enough to become sensitive to the presence of the substrate and that the confinement has to occur along the propulsion direction for it to influence the motion direction of the swimmer. With our work we provide an experimental demonstration that the shape of a microswimmer can be used to tune its self-propulsion direction. Our experiments also provide a new perspective on the as-of-yet still debated propulsion mechanism of Pt-catalysis driven microparticles, as the direction reversal can only be explained by a combination of the shape and the propulsion mechanism. 
\section*{Materials and Methods}

\subsection*{Reagents}
Fused silica substrates and IP-Dip photoresist were purchased from Nanoscribe GmbH. The platinum-palladium target (Pt/Pd 80:20) was acquired from Micro to Nano. Propylene glycol methylether acrylate (PGMEA, ReagentPlus \(\geq\) 99.5\%) was purchased from Sigma Aldrich and isopropanol (IPA) was obtained from VWR chemicals. Hydrogen peroxide ($H_2O_2$, 35 wt\% solution in water, stabilized) was purchased from Acros Organics. Water was filtered with a MilliQ MilliPore system (resistivity \(\geq\) 18 M\(\Omega\).cm). Unless otherwise noted, all chemicals were of analytical or reagent grade purity and were used as received from commercial sources.

\subsection*{Particle Fabrication}
All colloidal particles were 3D printed on a Nanoscribe Photonic Professional GT microprinter equipped with a 63x oil-immersion objective (Zeiss, NA = 1.4). We have described the printing routine in details in our previous work \cite{Doherty2020}. The spheres have a diameter of 7 $\mu$m. discs and tori were designed with a diameter of 10 $\mu$m and a thickness of 2 $\mu$m. The crescent structure corresponds to a half-tori with identical shape parameters. The rods are 10 $\mu$m long and 2 $\mu$m thick. Discs, tori, and crescents were printed standing up while rods were printed lying down on the substrate. All particles were printed on fused-silica substrates using IP-Dip as photoresist. The prints were developed in PGMEA for 30 min, followed by 2 min in IPA. Particles were made active by half coating them with a 5nm thick platinum-palladium film (Pt/Pd 80:20) using a Cressington 208HR sputter coater and argon as the inert gas in the sputtering chamber. Only afterwards the Janus microstructures were removed from the substrate by sonication in water, as described in ref. \cite{Riedel2024}. 
\\
\\
For active silica-coated particles a 50 nm thick silica film was applied after developing but before sputter coating with platinum-palladium. For this, a high vacuum Leybold-Heraeus Z400 sputter-coater with custom modifications was used to obtain a coating of all exposed particle surfaces. The target was positioned close to the sample and sputtering was performed at $\approx 10^{-5}$ mbar (Ar @ 54 sccm, Direct Current Electrde Positive (DCP): 1kV) which yields a robust layer that covers most of the particle.

\subsection*{Propulsion of Active Particles}
Pt-coated colloidal particles were suspended in freshly prepared aqueous hydrogen peroxide solution and placed in a sample holder ($\varnothing$ = 8 mm) with an untreated borosilicate glass coverslip (VWR, 25 mm, No. 1) as substrate. To prevent evaporation, the sample was covered with a second coverslip. The self-propulsion of these microswimmers arises from the catalytic decomposition of $H_2O_2$ at the Pt/Pd cap \cite{Ebbens2014, Howse2007}.

\subsection*{Data Acquisition and Analysis}
The active particles were observed under a Nikon Eclipse Ti-E bright-field light microscope with a Plan Apo $\lambda$ 20x long working distance objective (NA = 0.75). 30s long videos with a frame rate of 20 fps were recorded to obtain the average particle velocity for all individual shapes. 
\\
The particles were tracked by applying the Canny Edge detection algorithm to each frame to generate a mask from which the particle center of mass was obtained. Individual particle velocities $v$ and diffusion coefficients $\mathcal{D}$ were determined by fitting the mean squared displacements (MSDs) with $\Delta r^2 = 4\mathcal{D}\Delta t + v^2 \Delta t^2$ up to lag times much smaller than the rotational timescale for a sphere ($1.25~s \ll 50~s$). We here use the fit equation originally developed for spherical active particles \cite{Howse2007,Bechinger2016} as an approximation also for non-spherical particles. This yielded consistent and reliable results, albeit without subpixel resolution.  

\subsection*{z-Stack in Confocal Mode}
Z-stacked confocal images were captured with a Nikon Eclipse Ti microscope equipped with an A1R confocal scanning head and a 60× (NA = 1.2) water immersion objective. A z-stack image sequence consisted of multiple images recorded at different focal heights. A MCL Nano-Drive piezo was used to acquire high speed z-stack images for active 10 \si{\micro\meter} sized crescents in 1wt\% and 10wt\% H$_2$O$_2$ as well as for a 7 \si{\micro\meter} crescent in 1wt\% H$_2$O$_2$. The scanning was executed from bottom to top with a scan time of 0.6 sec per z-stack image.

\subsection*{Zeta-potential Measurements}
Zeta potential measurements were performed with a Malvern Zetasizer Nano ZS which determines the electrophoretic mobility of colloidal particles moving toward an electrode of opposite charge using a technique called Laser doppler velocimetry. The zeta potential is obtained from the electrophoretic mobility by applying the Henry equation, which is incorporated in the Zetasizer software. Two IP-Dip emulsions in water were prepared and cured under UV irradiation to obtain photoresist particles which were then used for zeta-potential measurements. One sample was cured for 2h (blue curve) and the other sample was cured over night (red curve), see Fig.\ref{Fig_6}b. The zeta potential of the bare particles was measured in aqueous $H_2O_2$ solution of 1 wt\%, 2 wt\%, 3 wt\%, 4 wt\%, 5 wt\%, 6 wt\%, 7 wt\%, 8 wt\% and 10 wt\%. 3D printed particles were not directly used for zeta potential measurements due to the long printing times required to obtain enough particles. The pH of all $H_2O_2$ solutions was measured with a pH110 pH meter from VWR. Fig.\ref{Fig_6}b shows the zeta-potential for both batches of bare IP-Dip particles. The particles cured for 2h (blue curve) were most probably not fully polymerized. The zeta potential remains negative over the entire pH range for both samples. 

\subsection*{COMSOL simulations}
The species distribution around the particles is simulated using COMSOL Multiphysics with a two-dimensional model incorporating the Transport of Diluted Species (tds) module. We simulate particles with both flat disk and spherical geometries, located either in bulk solution or near a solid substrate (300 nm above the substrate). The simulation domain is a square region of water measuring $200\,\mu\text{m} \times 200\,\mu\text{m}$. As a demonstration, we used protons ($\mathrm{H}^+$) as diffusing species. Only diffusion is considered for simplification. The diffusion coefficient of $\mathrm{H}^+$ is set to $D = 9.31 \times 10^{-9}\,\mathrm{m}^2/\mathrm{s}$, with an initial concentration of $2 \times 10^{-3}\,\mathrm{mol/m}^3$, based on the pH of 5.7 resulting from the dissolution of CO$_2$ in Milli-Q water. A flux of magnitude $J_{0,\mathrm{H}^+} = 1 \times 10^{-6}\,\mathrm{mol}/(\mathrm{m}^2 \cdot \mathrm{s})$ and is applied to the active sides of the particles and $\frac{1}{3} J_{0,\mathrm{H}^+}$ to the flat sides with reduced activity. The inert sides of the particles and the substrate are assigned a no-flux boundary condition. The outer boundaries of the water domain are maintained at the initial concentration, effectively mimicking an infinite bulk environment. A super fine mesh is applied to regions near the particles and substrate, built on top of a free triangular mesh.

\subsection*{Associated content}

\subsubsection*{Preprint} An initial draft of this work has been uploaded to the arXiv preprint server.  

\subsection*{Acknowledgments}
This work was supported by the Dutch Research Council (NWO/OCW), as part of the Vidi scheme (VI.Vidi.193.069, D.J.K.). We thank Joost de Graaf, Christina Kurzthaler and Ursy Makanga for useful discussions and Rachel Doherty for support with 3D printing and EDX measurements.
\textbf{Data availability}
Source data are provided with this paper and can be found through the 4TU public data repository under the link: 
\\
\\
doi: \url{https://doi.org/10.4121/d9fe4b8a-7276-452b-9e5d-1d51c1d093f5}
\\
\\
\textbf{Author contribution statement} 
DJK and SR conceived the work. SR carried out the experiments. MW performed the simulations. All authors discussed and interpreted the results and contributed to writing the final manuscript. 
\\
\\
\\
Daniela Kraft (Corresponding Author), Kraft@physics.leidenuniv.nl
\\
Solenn Riedel (Contributing Author 1), Riedel@physics.leidenuniv.nl
\\
Mengshi Wei (Contributing Author 2), MWei@physics.leidenuniv.nl
\\
\\
\textbf{Competing interests}
All authors declare no competing financial or non-financial interests. 
\\

\bibliography{References}
\newpage
\begin{center}
\title{{\Large\bf Supplementary Information:\\
Shape-dependent direction reversal in anisotropic catalytic microswimmers}}

\author{Solenn Riedel}
\author{Mengshi Wei}
\author{Daniela J. Kraft}
\end{center}

\maketitle

\section*{Particle design}
\begin{figure}[h!]
\centering
\includegraphics[width=0.8\textwidth]{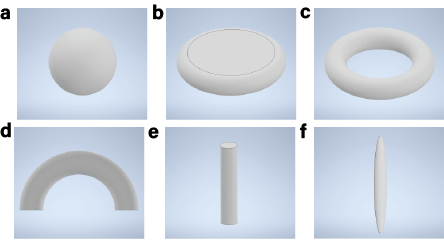}
\caption{CAD design for 7 \si{\micro\meter} spheres, 10 \si{\micro\meter} disks, 10 \si{\micro\meter} tori, 10 \si{\micro\meter} crescents, 10 \si{\micro\meter} rods and 10 \si{\micro\meter} ellipsoid.}
\label{fig:diff_consts}
\end{figure} 
\newpage\section*{Centered particle trajectories for crescents at different fuel concentrations}
\begin{figure}[h!]
\centering
\includegraphics[width=1\textwidth]{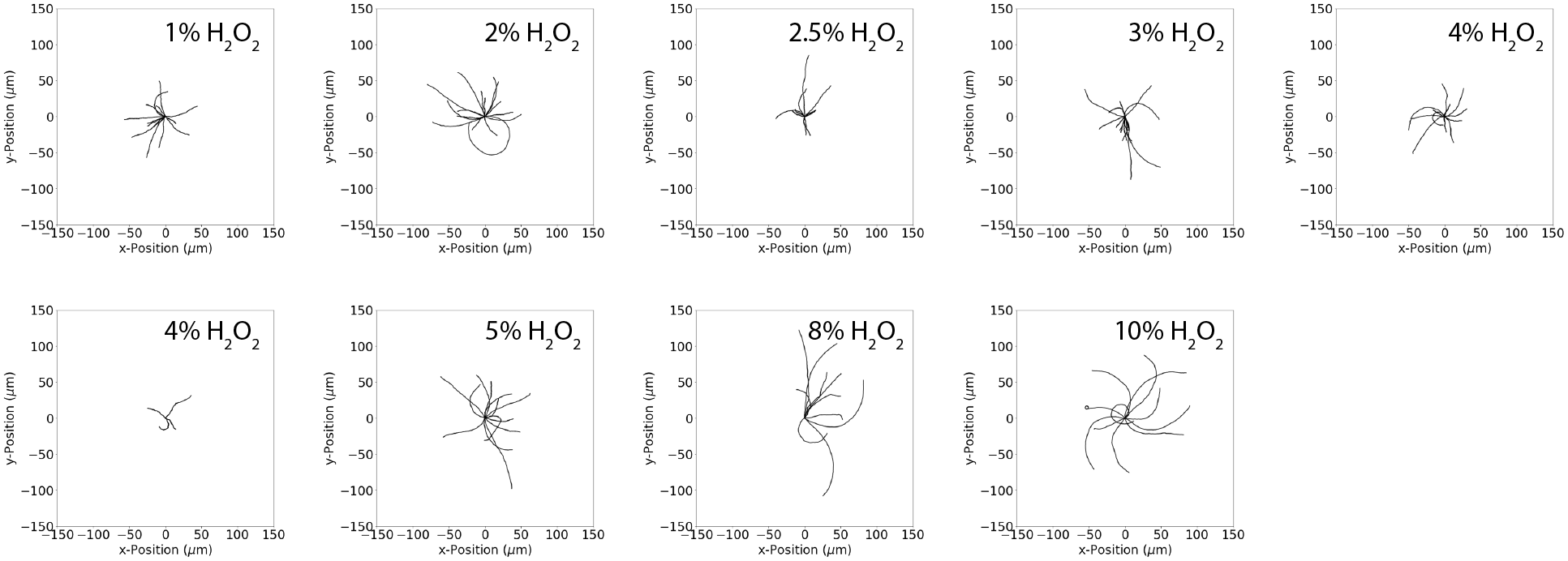}
\caption{30 sec trajectories for crescents at different H$_2$O$_2$ concentrations. Active crescents are swimming inert-side leading (top row) and catalytic-side leading (bottom row) depending on the fuel concentration. The direction reversal occurs at 4 wt\% H$_2$O$_2$ and both swimming directions are observed at this transition concentration. The average particle velocity extracted from this data can be found in Fig.2 d.}
\label{fig:trajectories}
\end{figure} 

\section*{Diffusion constants at different fuel concentrations for all shapes mentioned in this work}
\begin{figure}[h!]
\centering
\includegraphics[width=1\textwidth]{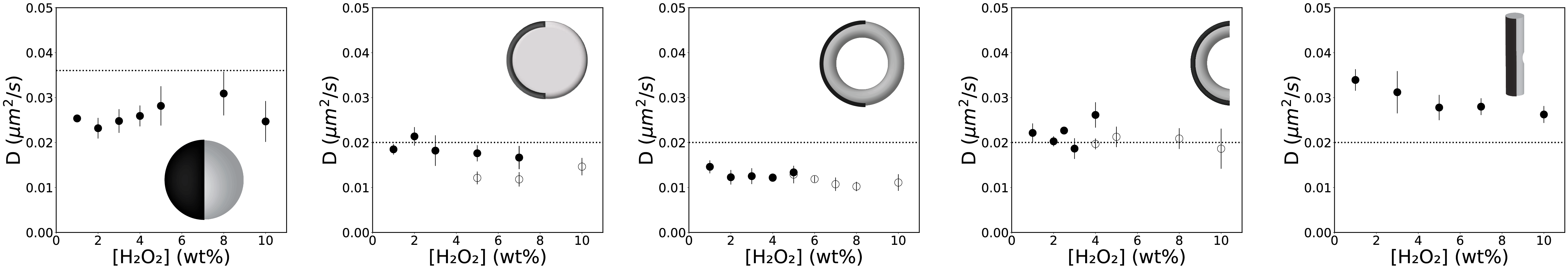}
\caption{Height-dependent diffusion constant $D$ for 7 \si{\micro\meter} spheres, 10 \si{\micro\meter} disks, 10 \si{\micro\meter} tori, 10 \si{\micro\meter} crescents and 10 \si{\micro\meter} rods in different fuel concentrations. The dotted line indicates the expectation value for an equally sized sphere close to the substrate with a radius of 3.5 or 5 \si{\micro\meter}, respectively \cite{Ketzetzi2020}. Plotted points are median values and error bars represent first and third quartiles. Black markers correspond to inert-side leading particles and white markers correspond to catalytic-side leading particles}
\label{fig:diff_consts}
\end{figure} 
\newpage
\section*{Flip of self-propelled crescents}
\begin{figure*}[h]
\centering
\includegraphics[width=1.0\textwidth]{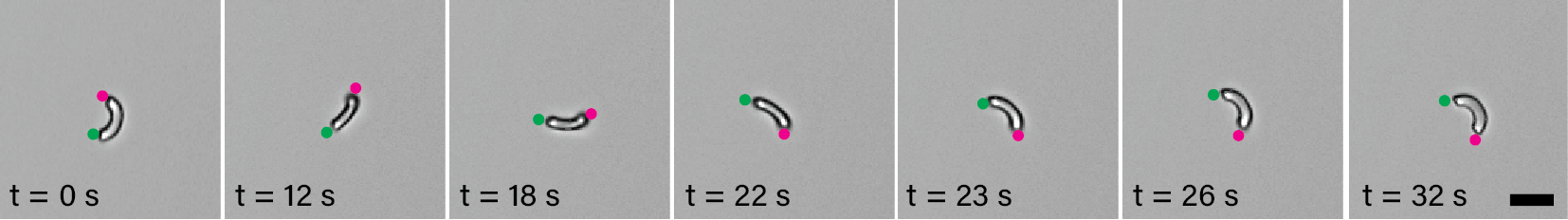}
\caption{\textbf{Flip of a SiO$_2$-coated crescent.} At a fuel concentrations corresponding to the transition region, we occasionally observe particles that flip. This can come with a change in swimming direction but this is not always the case. Here the crescent moves inert-side forward before and after the flip. The two crescent feet are color-coded to guide the eye and help the reader to see the flip. The scale bar is 10 \si{\micro\meter}.}
\label{SI_1}
\end{figure*}
\begin{figure*}[h]
\centering
\includegraphics[width=1.0\textwidth]{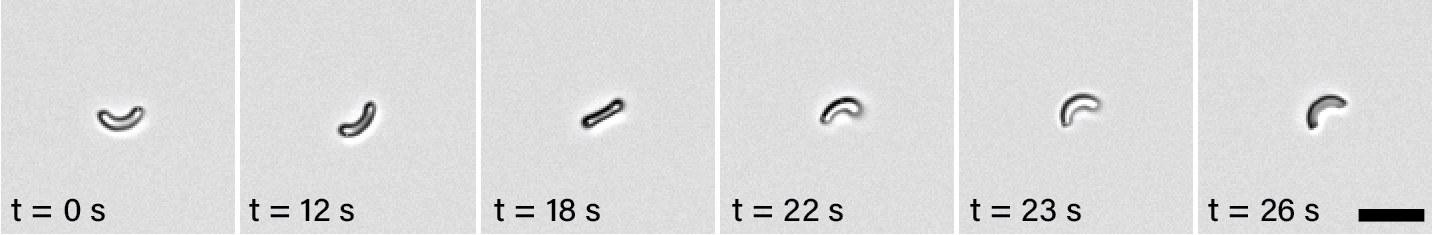}
\caption{\textbf{Flip of a 7 \si{\micro\meter} sized crescent.} Smaller crescents are also occasionally observed to flip at fuel concentrations close to the transition region. However, a change in swimming direction was never observed for flipping 7 \si{\micro\meter} crescents. The scale bar is 10 \si{\micro\meter}.}
\label{SI_2}
\end{figure*}
\newpage
\section*{Confocal z-stack imaging of 7 \si{\micro\meter} crescents}
\begin{figure*}[h]
\centering
\includegraphics[width=1.0\textwidth]{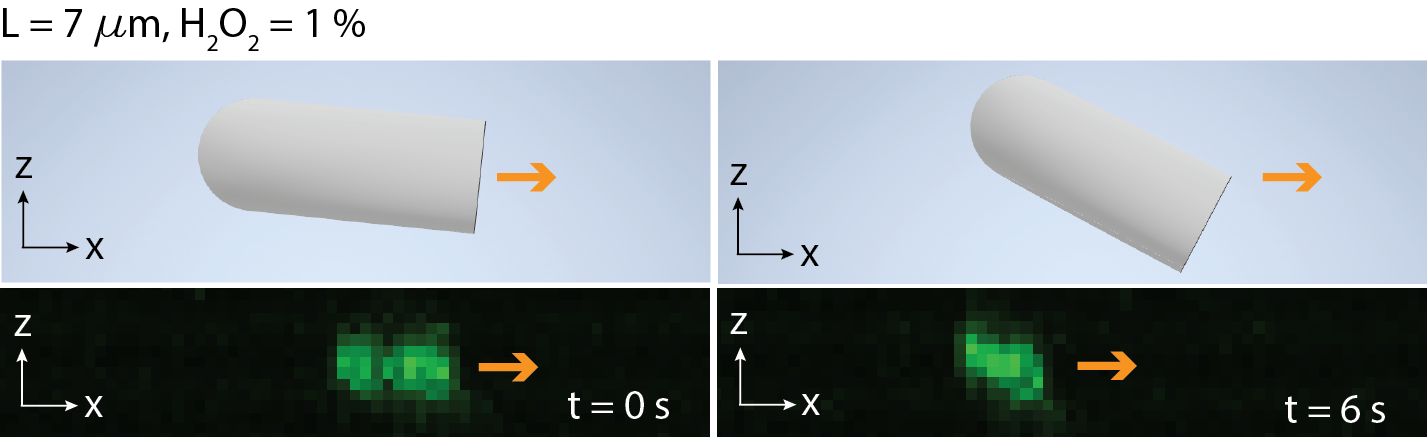}
\caption{\textbf{Orientation of a 7 \si{\micro\meter} crescents in 1wt\% H$_2$O$_2$, side view.} Projection of a confocal z-stack in the zx-plane and schematic illustrating the crescent orientation. The orientation of our smaller crescents is not stable and we see it fluctuate between a parallel (left) and a tilted (right, taken from main text) position.}
\label{SI_3}
\end{figure*}
\begin{figure*} [h]
\centering
\includegraphics[width=1.0\textwidth]{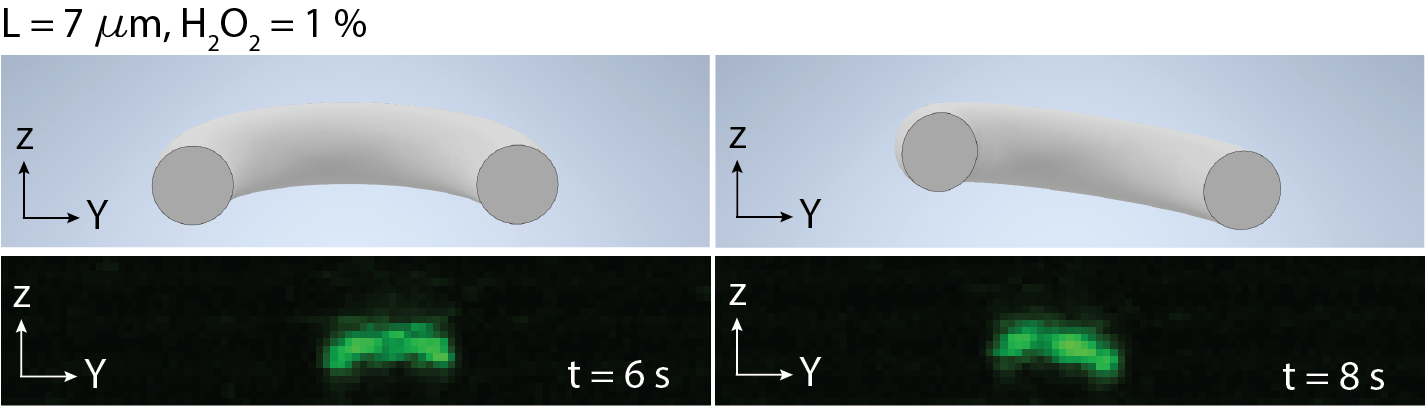}
\caption{\textbf{Orientation of a 7 \si{\micro\meter} crescents in 1wt\% H$_2$O$_2$, front view.} Projection of a confocal z-stack in the zy-plane and schematic illustrating the crescent orientation. An occasional tilt where both legs are not equally close to the substrate is also observed for smaller crescents in low fuel.}
\label{SI_4}
\end{figure*}
\newpage
\section{EDX measurements}
Energy-dispersive X-ray spectroscopy was used to identify the presence of oxide (PtO) the Pt/Pd metal target used in the sputter coating process. The sample was mounted on a scanning electron microscope (SEM) equipped with an integrated EDX detector and analyzed under high vacuum conditions. A general spectrum was acquired from a selected area of the sample surface to assess the overall elemental composition. Elemental quantification was performed using the Pathfinder software and spectral data revealed characteristic peaks corresponding to platinum (Pt), palladium (Pd), and oxygen (O). In the case of the target used for Fig. 3a, we expect the catalytic film sputter coated in Ar to show no PtO contamination if the elemental analysis of the target doesn´t show the presence of oxygen.
\begin{figure*}[h]
\centering
\includegraphics[width=1.0\textwidth]{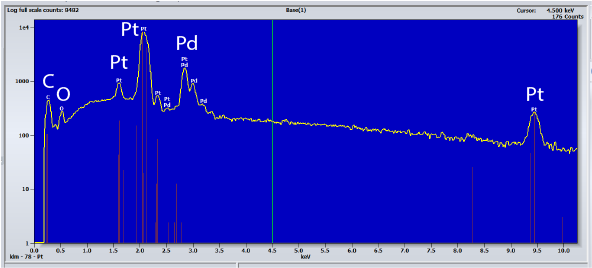}
\caption{\textbf{EDX spectrum of main target side that gets exposed to the plasma during sputter coating.} Next to the Pt and Pd peaks, a non-negligible oxygen peak is visible which confirms contamination with PtO.}
\label{SI_EDX_B}
\end{figure*}
\begin{figure*}[h]
\centering
\includegraphics[width=1.0\textwidth]{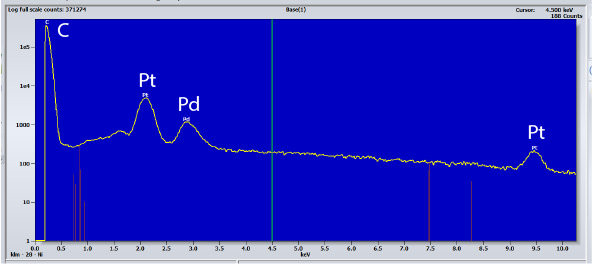}
\caption{\textbf{EDX spectrum of target used to prepare crescents from Fig.3a.} No oxygen peak is visible and a contamination of the catalytic cap material with PtO can thus be excluded.}
\label{SI_EDS_A}
\end{figure*}
\newpage
\section*{Water contact Angle of IP-Dip}
\begin{figure*}[h]
\centering
\includegraphics[width=0.8\textwidth]{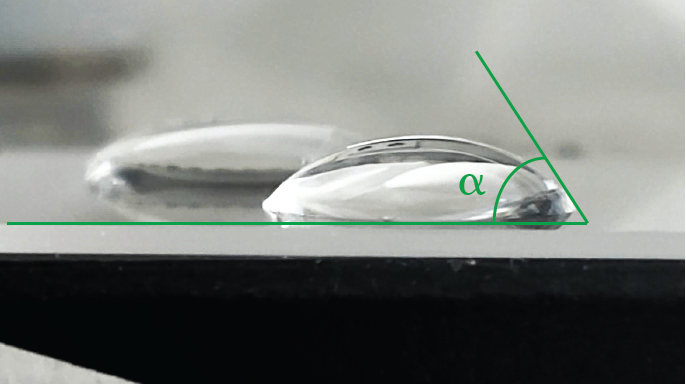}
\caption{\textbf{Water contact angle of IP-Dip} The commercial photoresist IP-Dip (Nanoscribe GmbH) was spinn-coated on a glass substrate and cured under UV for 90 hours. The water contact angle (CA) is ca. $\sim$\ang{50}.}
\label{SI_6}
\end{figure*}
\newpage
\section*{Description of the Videos}
\noindent Supplementary Video 1: 3D-printed active sphere in 1 wt\% H$_2$O$_2$: \\
Sphere self-propelling in a 1\% aqueous H$_2$O$_2$ solution. The particle moves inert-side leading. The original frame rate is 20 fps and the speed was enhanced 10x. 
\\
\\
Supplementary Video 2: 3D-printed active sphere in 10 wt\% H$_2$O$_2$: \\
Sphere self-propelling in a 10\% aqueous H$_2$O$_2$ solution. The particle moves inert-side leading. The original frame rate is 20 fps and the speed was enhanced 10x.
\\
\\
Supplementary Video 3:  3D-printed disk in 1 wt\% H$_2$O$_2$:\\
Disk self-propelling in a 1\% aqueous H$_2$O$_2$ solution. The particle moves inert-side leading. The original frame rate is 20 fps and the speed was enhanced 10x.
\\
\\
Supplementary Video 4:  3D-printed disk in 10 wt\% H$_2$O$_2$:\\
Disk self-propelling in a 10\% aqueous H$_2$O$_2$ solution. The particle moves catalytic-side leading. The original frame rate is 20 fps and the speed was enhanced 10x.
\\
\\
Supplementary Video 5:  3D-printed torus in 1 wt\% H$_2$O$_2$:\\
Torus self-propelling in a 1\% aqueous H$_2$O$_2$ solution. The particle moves inert-side leading. The original frame rate is 20 fps and the speed was enhanced 10x.
\\
\\
Supplementary Video 6:  3D-printed torus in 10 wt\% H$_2$O$_2$:\\
Torus self-propelling in a 10\% aqueous H$_2$O$_2$ solution. The particle moves catalytic-side leading. The original frame rate is 20 fps and the speed was enhanced 10x.
\\
\\
Supplementary Video 7: 3D-printed bent rod in 1 wt\%  H$_2$O$_2$: \\
Bent rod self-propelling in a 1\% aqueous H$_2$O$_2$ solution. The particle moves inert-side leading. The original frame rate is 20 fps and the speed was enhanced 10x.
\\
\\
Supplementary Video 8: 3D-printed bent rod in 5 wt\% H$_2$O$_2$:\\
Bent rod self-propelling in a 5\% aqueous H$_2$O$_2$ solution. The particle moves catalytic-side leading. The original frame rate is 20 fps and the speed was enhanced 10x.
\\
\\
Supplementary Video 9: 3D-printed straight rod in 1 wt\%  H$_2$O$_2$: \\
Rod self-propelling in a 1\% aqueous H$_2$O$_2$ solution. The particle moves inert-side leading. The original frame rate is 20 fps and the speed was enhanced 10x.
\\
\\
Supplementary Video 10: 3D-printed straigth rod in 10 wt\%  H$_2$O$_2$: \\
Rod self-propelling in a 10\% aqueous H$_2$O$_2$ solution. The particle moves inert-side leading. The original frame rate is 20 fps and the speed was enhanced 10x.
\\
\\
\end{document}